\documentclass[conference]{IEEEtran}
\usepackage{ifpdf}
\usepackage{cite}
\usepackage{tikz}
\usepackage{lipsum}
\usepackage{mathtools}
\usepackage{cuted}
\usepackage[left=0.63in,right=0.63in,top=0.67in,bottom=0.63in]{geometry}
\usetikzlibrary{decorations, decorations.text,}
\usepackage{float}
\usepackage{amsmath}
\usepackage{amsthm}
\usepackage{mathtools, cuted}
\usepackage{array}
\usepackage[caption=false,font=footnotesize]{subfig}
\usepackage{textcomp}
\usepackage{url}
\usepackage{paralist,esint,amssymb,booktabs,cases,bm,galois}

\newcommand\Se[1]{\mathcal{#1}}
\newcommand\Db[1]{\mathbb{#1}}

\newcommand\SB[1]{\left(#1\right)}

\newcommand{\RN}[1]{\textup{\uppercase\expandafter{\romannumeral#1}}}
\usepackage{cleveref}
\usetikzlibrary{calc}
\usetikzlibrary{shadings}
\newtheorem{theo}{Theorem}

\newtheorem{exam}{Example}

\newtheorem{rem}{Remark}

\newtheorem{claim}{Claim}

\usepackage{algorithm}
\usepackage[noend]{algpseudocode}
\makeatletter
\def\BState{\State\hskip-\ALG@thistlm}
\makeatother

\errorcontextlines\maxdimen
\makeatletter
\newcommand*{\algrule}[1][\algorithmicindent]{\makebox[#1][l]{\hspace*{.5em}\vrule height 0.9 \baselineskip depth 0.3\baselineskip}}%

\newcount\ALG@printindent@tempcnta
\def\ALG@printindent{%
    \ifnum \theALG@nested>0
        \ifx\ALG@text\ALG@x@notext
            \addvspace{0pt}
        \else
            \unskip
            \ALG@printindent@tempcnta=1
            \loop
                \algrule[\csname ALG@ind@\the\ALG@printindent@tempcnta\endcsname]%
                \advance \ALG@printindent@tempcnta 1
            \ifnum \ALG@printindent@tempcnta<\numexpr\theALG@nested+1\relax
            \repeat
        \fi
    \fi
    }%
\usepackage{etoolbox}
\patchcmd{\ALG@doentity}{\noindent\hskip\ALG@tlm}{\ALG@printindent}{}{\errmessage{failed to patch}}
\makeatother



\begin{document}

\bstctlcite{IEEEexample:BSTcontrol}
\title{{Efficient Information Reconciliation for Energy-Time Entanglement Quantum Key Distribution\vspace{-6pt}}}
\author{\IEEEauthorblockN{Siyi Yang, Murat Can Sarihan, Kai-Chi Chang, Chee Wei Wong, and Lara Dolecek}
\IEEEauthorblockA{Electrical and Computer Engineering Department, University of California, Los Angeles, Los Angeles, CA 90095 USA\\
siyiyang@ucla.edu, mcansarihan@ucla.edu, uclakcchang@ucla.edu, cheewei.wong@ucla.edu, and dolecek@ee.ucla.edu\vspace{-6pt}
}}
\maketitle

\begin{abstract} Graph based codes such as low density parity check (LDPC) codes have been shown promising for the information reconciliation phase in quantum key distribution (QKD). However, existing graph coding schemes have not fully utilized the properties of the QKD channel. In this work, we first investigate the channel statistics for discrete variable (DV) QKD based on energy-time entangled photons. We then establish a so-called balanced modulation scheme that is promising for this channel. Based on the modulation, we propose a joint local-global graph coding scheme that is expected to achieve good error-correction performance.

Key words: quantum key distribution, information reconciliation, graph based codes, joint local-global graph based coding.
\end{abstract}

\IEEEpeerreviewmaketitle
\section{Introduction}
\label{section: introduction} 
Unconditional security in data communication is currently being deployed in commercial applications. Nonetheless, it faces a number of important challenges such as secret key rate, distance, size, cost and practical security. Quantum communications promise to deliver unprecedented levels of security, and to fundamentally change how we transmit and process sensitive data. An important phase of quantum communications is the Quantum Key Distribution (QKD). 

QKD offers a physically secure way for effectively sharing an encryption key over a quantum communication channel observed by an eavesdropper. QKD protocols based on energy-time entangled photons have been studied due to their ability to provide a higher secure key rate by carrying multiple bits per entangled photon pair and resulting in higher information efficiency
per channel \cite{Zhong2015a}. Energy-time entangled photons as a secure QKD platform are theoretically proven to offer physically unconditional security for key generation \cite{Zhang2013}. Information reconciliation is an important phase in QKD protocols, where error-correction codes are applied to ensure that the two parties share an identical random sequence that is used as private key for one-time-pad encrypted communication.

Graph based codes such as irregular LDPC and spatially-coupled (SC) LDPC codes have been considered for information reconciliation \cite{zhou2013layered,jiang2018high}. In particular, in \cite{zhou2013layered}, layered coding based on splitting the joint channel into channels for each bit layer, according to the chain rule of mutual information, offers a key rate higher than that of joint coding (treating the concatenation of bits as the bit-stream). Avoiding high latency to keep pace with the speed at which raw key is generated has become increasingly more important, see, e.g.,\cite{martinez2013key,Wehnereaam9288}. Layered coding is computationally challenging in the finite-length regime, especially when the number of bits per symbol increases, due to the dependencies between layers and the unexplored discrepancy between the asymptotic analysis and the finite-length code construction \cite{olmos2015scaling}. Non-binary graph codes, e.g., NB-LDPC codes, have been shown to offer better performance than binary codes at the same rate. However, the complexity of iterative decoders for NB-LDPC codes on a Galois field $\textup{GF}(q)$ is typically of an order of magnitude $O(q\log q)$ higher than that for binary LDPC codes, which leads to a significantly higher latency that prohibits them from practical implementations \cite{declercq2007decoding}. In order to possess both reliability and low complexity, we therefore propose a so-called ``joint local-global coding'' scheme that combines local binary graph codes and global non-binary graph codes together. This coding scheme not only allows a low latency parallel implementation of iterative decoding, but also provide significant flexibility in the code design.  

\section{System Model}
\label{section: system model}
\subsection{Information Reconciliation}
\label{subsection: info reconciliation}

\begin{figure}
  \centering
  \includegraphics[width=0.9\linewidth]{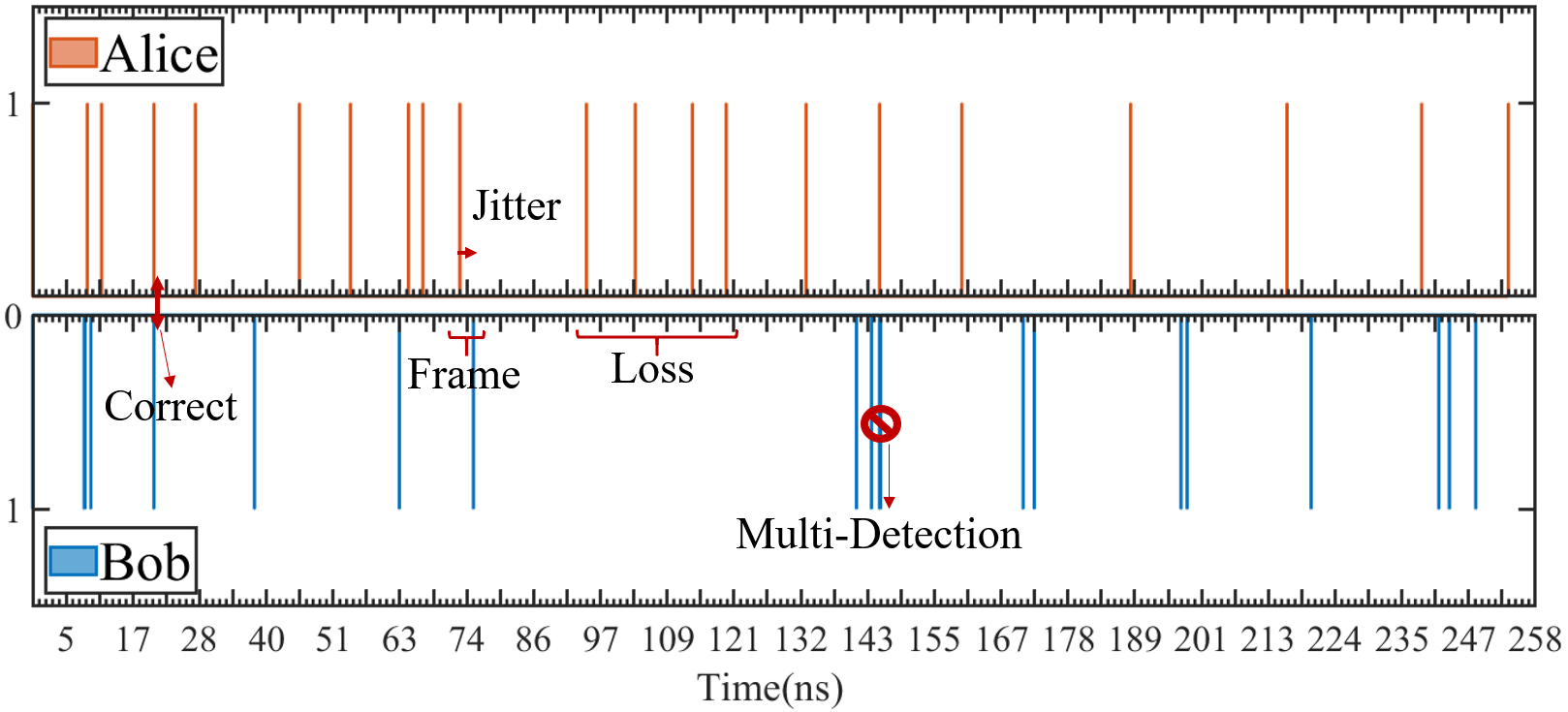}
  \caption{A data stream obtained from a time-bin QKD test bed where $M=6$. The common sources of error for energy-time entanglement based time-bin protocols are in two groups; either the local errors caused by timing jitter errors, or global errors caused by photon losses, multi-pair detections and accidental coincidences.}
  \label{fig:QKD}
\end{figure}
In the time-bin protocol\cite{Zhong2015a}, energy-time entangled photon pairs are generated by Alice or a third party: one photon out of this pair is delivered to Bob over a quantum channel while the other one is retained by Alice. The information is encoded to photon arrival time that should be identical for entangled photons; hence Alice and Bob record the arrival time of the
detected photons with a synchronized clock. The time domain is divided into non-overlapping frames and each frame consists of $2^M$ bins of an identical length, thus each arrival time is represented by a $M$-bit symbol on $\textup{GF}(2^M)$. Detections in a frame where both Alice and Bob have exactly one detection are regarded effective and others are discarded.

The sequences that Alice and Bob receive are supposed to be identical if the environment is perfect. However, in practical implementation, they suffer from time jitter in detector electronics, which results in discrepancies between the recorded sequences. Furthermore, the loss of one of the photons in each pair or accidental concurrent detections for a given frame may also pose an error in the data stream. A raw data-stream from our testbed is shown in Fig.~\ref{fig:QKD} and the cases affecting reliable communication are annotated.

From the previous discussion, the raw sequence recorded by Bob is essentially a noisy version of that recorded by Alice. Therefore, Bob is able to ``decode'' the sequence of Alice, which later on will be used as the private key for secure communication between Alice and Bob, through error correction coding. This process is known as ``information conciliation''.

\subsection{Channel Characterization}
\label{subsection: channel char}
In the information reconciliation phase, the sequences obtained by Alice and Bob are $X^N$ and $Y^N$, respectively, where $X,Y\in \textup{GF}(2^M)$ and $N$ is the sequence length. A syndrome $R=\bold{H}X$ is sent to Bob so that he is able to recover $X$ from $R$ and $Y$, where $\bold{H}$ is the parity check matrix of a code over $\textup{GF}(2^M)$. 

\begin{figure}
  \centering
  \includegraphics[width=0.7\linewidth]{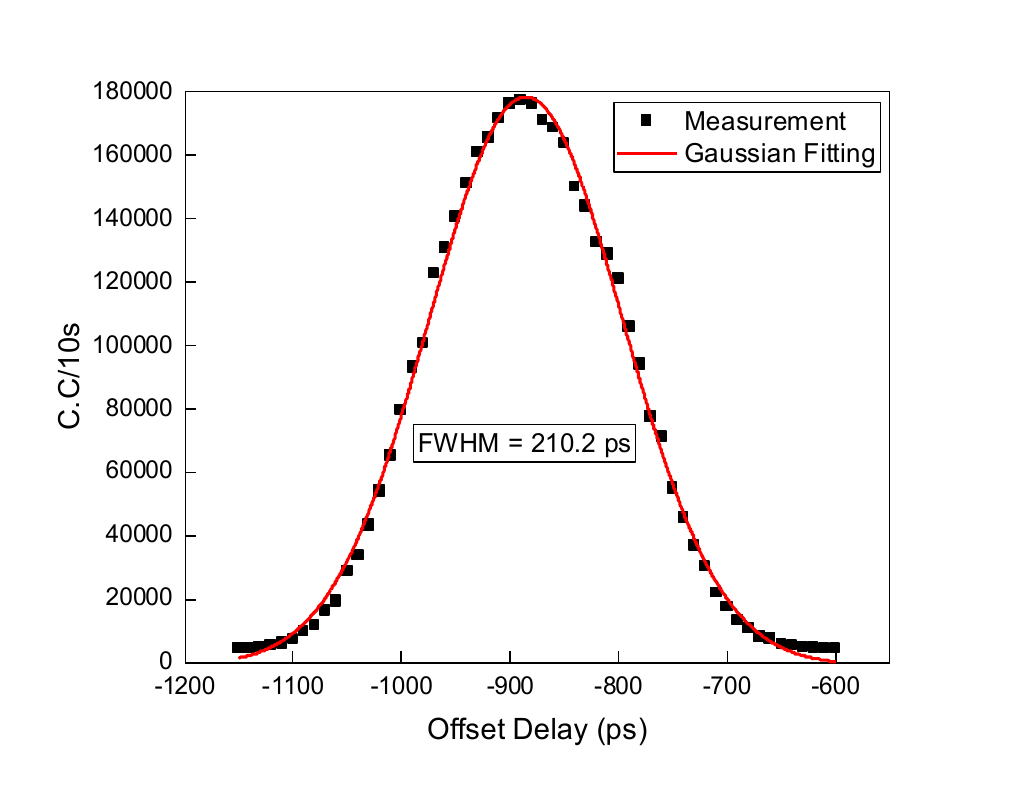}
  \caption{Channel characterization for single-channel time-bin encoding. Uncertainty in coincidence timestamp for the entangled pair, which is due to the convolved timing jitter of the detector.}
  \label{fig:channel}
\end{figure}

We observe that the channel fits into a mixture of Gaussian distribution and uniform distribution, as is shown in Fig.~\ref{fig:channel}, which is simplified as the following transitional probability: 
\begin{equation}
\label{eqn: channelmodel}
p_{X|Y}(x|y)=ce^{-\frac{|x-y|^2}{\sigma}}+\beta,
\end{equation}
where $\sigma$ and $\beta$ are parameters determining the strength of the local Gaussian channel and the global uniform channel, respectively, and $c$ is a normalization constant. This is also the local-global channel described in \cite{zhou2013layered}.

A closer inspection of errors suggests that the local channel and the global channel are caused by different error sources, referred to as the local errors and the global errors, respectively. Local errors originate from timing jitters and synchronization errors in photon detection. Under local errors, the two photons from the same pair fall into different but close enough bins within the same frame. Global errors are caused by channel losses, and accidental concurrent detections of stray photons not associated with any entangled photon pairs. Within-frame multi-photon detections also contribute to global errors. Experimental results show that there exists a nontrivial dependency between the bin width vs. the fraction of local errors and the global errors, as well as the number of bits per symbol. For each selected field size, a bin width is chosen to optimize the raw symbol error rate.


\section{Modulation and Coding Schemes}
\label{section: modulation and coding schemes}
In \Cref{subsection: channel char}, we have discussed the QKD channel model. In this section, we present a so-called ``joint local-global coding'' scheme for efficient error correction of local and global errors in this channel. 

\subsection{Balanced Modulation}
In this subsection, we investigate the modulation scheme, the so-called ``balanced modulation'', which is appropriate for the proposed joint coding scheme presented in the next subsection. 
\begin{claim} A good modulation scheme $A$ for the channel discussed in \Cref{subsection: channel char} maps ${0,1,\cdots,2^M-1}$ to $\textup{GF}(2^M)$, and is expected to possess the following properties and is referred to as a balanced modulation.
\begin{enumerate}
\item The neighboring symbols have few distinct bits.
\item When considering the $l$-th level bit string in isolation, i.e., $\bold{s}_l=\SB{A(1)_l,A(2)_l,\cdots,A(2^M)_l}$, the number of runs within each $\bold{s}_l$ should be as small as possible.
\end{enumerate} 
\end{claim}
The first property is natural for channels possessing high locality. To motivate the second property, we consider the following example.

\begin{exam} When $M=4$, the Gray code-based modulation maps $0,1,\cdots,15$ to $0000$, $0001$, $0011$, $0010$, $0110$, $0111$, $0101$, $0100$, $1100$, $1101$, $1111$, $1110$, $1010$, $1011$, $1001$, $1000$, respectively. In the balanced distribution, $0,1,\cdots,15$ are mapped to $0000$, $0001$, $0011$, $0111$, $1111$, $1110$, $1100$, $1000$, $1001$, $1011$, $1010$, $0010$, $0110$, $0100$, $0101$, $1101$, respectively. Denote the error probability of the least significant bit (LSB) for these two modulations by $p^{\textup{G}}_e$ and $p^{\textup{B}}$, respectively, where $(1-p_0)$ denotes the raw symbol error rate. We see that $p^{\textup{G}}_e\sim\frac{1-p_0}{2}$ (even if the other bits are perfectly correct) and $p^{\textup{B}}_e\sim\frac{5}{8}p^{\textup{G}}_e$. Therefore, the FER contributed by the erroneous decoding of LSB in the Gray coded modulation is a bottleneck for bit-level encoding.
\end{exam}

Generally, a good $M$-bit balanced modulation should make the bit error probability on each bit close to $\frac{1-p_0}{M}$, which not only removes the burden of designing ultra low rate codes for the LSB in the Gray coded modulation, but also simplifies the code design since using the same constituent binary code for each bit-level in the joint local-global coding scheme (to be specified later) is sufficient for a code with good FER. 

\subsection{Joint Local-Global LDPC Codes}

\begin{figure}
  \centering
  \includegraphics[width=0.85\linewidth]{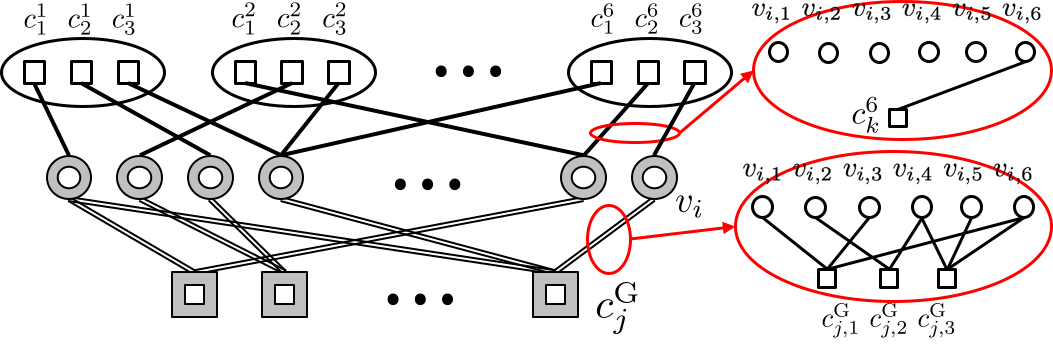}
  \caption{Joint local-global LDPC code. This picture presents a joint local-global LDPC code with $M=6$, $W=3$.}
  \label{fig:picture2}
\end{figure}

The Tanner graph of a joint local-global graph code is indicated by Fig.~\ref{fig:picture2}. 
 Each variable node (VN) $v_{i}$ has $M$ bits, represented by $v_i=(v_{i,1},v_{i,2},\dots,v_{i,M})$, $1\leq i\leq N$. There are two types of check nodes (CNs): the set of local CNs, denoted by $\Se{A}^{\textup{L}}$, and the set of global CNs, denoted by $\Se{A}^{\textup{G}}$. 

 The local CNs are further partitioned into $M$ groups, represented by $\Se{A}_l$, $1\leq l\leq M$, respectively. CNs in $\Se{A}_l$ only connect to the $l$-th bits of the VNs, and are denoted by $c^l_k$, $1\leq k\leq |\Se{A}_l|$. The edges from $\Se{A}_l$ is represented by an adjacent matrix $\bold{H}^{\textup{L}}_l$. 
 SC-LDPC codes, with their excellent waterfall performance, are adequate to be applied here\cite{hareedy2018spatially,esfahanizadeh2018finite}. 

 Each global CN in $\Se{A}^{\textup{G}}$, denoted by $c^{\textup{G}}_j$, $1\leq j\leq |\Se{A}^{\textup{G}}|$, is a compound CN consisting of $W$ CNs $c^{\textup{G}}_{j,w}$, $1\leq w\leq W$. Any compound edge connecting $c^{\textup{G}}_j$ and $v_i$ contains multiple edges connecting $v_{i,l}$ and $c^{\textup{G}}_{j,w}$, and is described by an adjacent matrix $\bold{H}_{i,j}\in \textup{GF}(2)^{W\times M}$. Each $\bold{H}_{i,j}$ is randomly chosen from a set $\Se{H}$ of elements from $\textup{GF}(2)^{W\times M}$. Note that $2\leq W\leq M$, and when $W=M$ and $\Se{A}^{\textup{L}}=\emptyset$, the global CN part in the joint local-global code degenerates to a non-binary LDPC code of $\textup{GF}(2^M)$. Smaller $W$ is expected to offer better code rate without a huge degradation of performance because of the local property of the channel.

\subsection{Joint Local-Global Belief Propagation}

The traditional belief propagation can be generalized to be applied to the decoding algorithm of our proposed code. Denote the sets of neighboring VNs of $c_j^l$ and $c_j$ by $\Se{N}_j^l$, $\Se{N}_j^{\textup{G}}$, respectively. Denote the sets of neighboring local CNs from $\Se{A}_l$ and global CNs of $v_i$ by $\Se{T}^{l}_{i}$ and $\Se{T}^{\textup{G}}_{i}$, respectively. Let $\Delta=Y-X$ over $\textup{GF}(2^M)$, then, $\Delta_i=Y_i-X_i$, $\Delta_{i,l}=Y_{i,l}-X_{i,l}$, where $1\leq i\leq N$, $1\leq l\leq M$. Denote the syndromes corresponding to check nodes $c^l_j$ and $c^{\textup{G}}_j$ in $S=\bold{H}Y-R=\bold{H}(Y-X)=\bold{H}\Delta$ by $s^l_k$ and $\bold{s}^{\textup{G}}_j$, respectively. 

The messages transmitted between VNs and global CNs are vectors of $2^M$ log-likelihood-ratios (LLRs). Let $L\left[t\right]$ denote the $t$'th component of vector $L$, $t\in \textup{GF}(2^M)$. Denote the messages transmitted between VN $v_i$ and local CN $c_j^l\in\Se{A}_l$ by $L_{v_i\to c_j^l}$ and $L_{c_j^l\to v_i}$, respectively. Denote the messages transmitted between VN $v_i$ and local CN $c_j^\textup{G}\in\Se{A}^{\textup{G}}$ by $L_{v_i\to c_j^\textup{G}}$ and $L_{c_j^\textup{G}\to v_i}$, respectively. In the remaining context, the superscript $(m)$ denotes the $m$-th round of iteration.

\begin{theo} The updating laws of joint local-global belief propagation algorithm are specified as follows:
\begin{equation*}\small
\begin{split}
&L_{v_i\to c_j^l}^{(m)}=f_{M,l}(L^{\textup{ch}}_i)+\sum\limits_{k\in\mathcal{T}^l_i\hspace{-1pt}\setminus\hspace{-1pt}\{c^l_j\}} L^{(m-1)}_{c^l_k\to v_i}+f_{M,l}\left(\sum\limits_{k\in\mathcal{T}_i^{\textup{G}}}L^{(m-1)}_{c_k^{\textup{G}}\to v_i} \right),\\
&L_{c_j^l\to v_i}^{(m)}=(1-2s^l_j)\mathcal{L}\left(\phi,\{ L_{v_{k}\to c^l_j} \}_{k\in \mathcal{N}_j^l\hspace{-1pt}\setminus\hspace{-1pt}\{i\}} \right),\\
&L_{v_i\to c_j^{\textup{G}}}^{(m)}\left[t\right]=L^{\textup{ch}}_{i,t}+\hspace{-9pt}\sum\limits_{k\in \mathcal{T}^{\textup{G}}_{i}\hspace{-1pt}\setminus\hspace{-1pt} \{c_j^{\textup{G}}\}}\hspace{-9pt} L^{(m-1)}_{ c_k^{\textup{G}}\to v_i}\left[t\right]-\sum\limits_{l=1}^{M}\Db{I}(t_l=1)\sum\limits_{k\in \mathcal{T}_i^l} L^{(m-1)}_{c_k^l\to v_i},\\
&L^{(m)}_{c_j^{\textup{G}}\to v_i}\left[t\right]=L^{'(m-1)}_{c_j^{\textup{G}}\to v_i}\left[s_j^{\textup{G}}+\bold{H}_{j,i}t\right]-L^{'(m-1)}_{c_j^{\textup{G}}\to v_i}\left[s_j^{\textup{G}}\right],
\end{split}
\end{equation*}
where $\phi(x)=-\log(\tanh(x/2))$, and $L_i^{\textup{ch}}$, $f_{M,l}$, $\mathcal{L}$, $L'_{c_j^{\textup{G}}\to v_i}$ are defined as follows:
\begin{equation*}\small
\begin{split}
&L_i^{\textup{ch}}\left[t\right]=\log \frac{P(\Delta_i=t|Y_i=y_i)}{P(\Delta_i=0|Y_i=y_i)},\\
&f_{M,l}(L)=\log \left(\sum\nolimits_{t_l=0} e^{L\left[t\right]}\right)-\log \left(\sum\nolimits_{t_l=1} e^{L\left[t\right]}\right),\\
&\mathcal{L}\left(f,\{a_i\}_{i\in I}\right)=\prod\nolimits_{i\in I}\textup{sgn}(a_i)f^{-1}\left(\sum\nolimits_{i\in I} f(\lvert a_i\rvert)\right) ,\\
&L'_{c_j^{\textup{G}}\to v_i}=\mathcal{F}_W^{-1}\comp \mathcal{L}\left(\log x, \big \{\mathcal{F}_W \comp h\left(\mathbf{H}_{j,k}, L_{v_k\to c^{\textup{G}}_j} \right)\big \}_{k\in\mathcal{N}^{\textup{G}}_j\hspace{-1pt}\setminus\hspace{-1pt}\{i\}}\right).
\end{split}
\end{equation*}
\end{theo}

\begin{rem} 
Here $\Se{F}_W$ refers to the Fast Fourier Transform (FFT) on Galois field $\textup{GF}(2^M)$, i.e., $\Se{F}_W(\mathbf{p})=\mathbf{H}^{\otimes M}\mathbf{p}$, where $\mathbf{p}\in \mathbb{R}^{2^M}$ and $\mathbf{H}$ is defined as follows:
\begin{equation*}
\mathbf{H}=\frac{1}{2}\left[\begin{array}{cc}
1&1\\
1&-1\\
\end{array}\right].
\end{equation*}
It is not hard to verify that for vectors $\mathbf{p}_1$, $\mathbf{p}_2$ representing probability distributions of two random variables $X_1$, $X_2$, over $GF(2^M)$, the probability distribution of $X_1+X_2$ is $\Se{F}_W^{-1}\left(\Se{F}_W(\mathbf{p}_1)\Se{F}_W(\mathbf{p}_2)\right)$.
\end{rem}

\begin{proof} Based on similar assumptions specified in \cite{richardson2001design}, i.e., the independence of constraints representing each neighboring CN of each VN conditioned on that VN, we obtain the following VN-to-CN updating laws,
\begin{equation*}\small
\begin{split}
&L_{v_i\to c_j^l}\\
=&\log \frac{P(\Delta_{i,l}=0|Y_i=y_i,\{\Se{T}^{l'}_{i}\hspace{-3pt}\setminus\hspace{-3pt}\{c_j^l\}\}_{l'=1}^{M},\Se{T}^{\textup{G}}_{i} )}{P(\Delta_{i,l}=1|Y_i=y_i,\{\Se{T}^{l'}_{i}\hspace{-3pt}\setminus\hspace{-3pt}\{j\}\}_{l'=1}^{M},\Se{T}^{\textup{G}}_{i}  )}\\
=&\log \frac{P(Y_i=y_i,\{\Se{T}^{l'}_{i}\hspace{-3pt}\setminus\hspace{-3pt}\{c_j^l\}\}_{l'=1}^{M},\Se{T}^{\textup{G}}_{i} |\Delta_{i,l}=0)}{P(Y_i=y_i,\{\Se{T}^{l'}_{i}\hspace{-3pt}\setminus\hspace{-3pt}\{c_j^l\}\}_{l'=1}^{M},\Se{T}^{\textup{G}}_{i} |\Delta_{i,l}=1)}\\
=&\log \frac{P(\Delta_{i,l}=0|Y_i=y_i)}{P(\Delta_{i,l}=1|Y_i=y_i)}+\log \frac{P(\{\Se{T}^{l'}_{i}\hspace{-3pt}\setminus\hspace{-3pt}\{c_j^l\}\}_{l'=1}^{M},\Se{T}^{\textup{G}}_{i} |\Delta_{i,l}=0)}{P(\{\Se{T}^{l'}_{i}\hspace{-3pt}\setminus\hspace{-3pt}\{c_j^l\}\}_{l'=1}^{M},\Se{T}^{\textup{G}}_{i} |\Delta_{i,l}=1)}\\
=&f_{M,l}(L^{\textup{ch}}_i)+\sum_{l'=1}^{M} \log \frac{P(\Se{T}^{l'}_{i}\hspace{-3pt}\setminus\hspace{-3pt}\{c^l_j\} |\Delta_{i,l}=0)}{P(\Se{T}^{l'}_{i}\hspace{-3pt}\setminus\hspace{-3pt}\{c^l_j\}  |\Delta_{i,l}=1)}+\log \frac{P(\Se{T}^{\textup{G}}_{i} |\Delta_{i,l}=0)}{P(\Se{T}^{\textup{G}}_{i} |\Delta_{i,l}=1)}\\
=&f_{M,l}(L^{\textup{ch}}_i)+\sum\limits_{k\in\Se{T}^l_i\hspace{-1pt}\setminus\hspace{-1pt}\{c^l_j\}} L_{c^l_k\to v_i}+\sum\limits_{l'\neq l}\sum\limits_{k\in\Se{T}^{l'}_i} \log \frac{P(c_k^{l'}|\Delta_{i,l}=0)}{P(c_k^{l'}|\Delta_{i,l}=1)}\\ 
&+ \log \frac{P(\Delta_{i,l}=0|\Se{T}^{\textup{G}}_{i} )}{P(\Delta_{i,l}=1|\Se{T}^{\textup{G}}_{i} )}\\
=&f_{M,l}(L^{\textup{ch}}_i)+\sum\limits_{k\in\Se{T}^l_i\hspace{-1pt}\setminus\hspace{-1pt}\{c^l_j\}} L_{c^l_k\to v_i}+f_{M,l}\left( \log \frac{P(\Delta_{i}|\Se{T}^{\textup{G}}_{i} )}{P(\Delta_{i}=0|\Se{T}^{\textup{G}}_{i} )}\right)\\
=&f_{M,l}(L^{\textup{ch}}_i)+\sum\limits_{k\in\Se{T}^l_i\hspace{-1pt}\setminus\hspace{-1pt}\{c^l_j\}} L_{c^l_k\to v_i}+f_{M,l}\left(\sum\limits_{k\in\Se{T}_i^{\textup{G}}}L_{c_k^{\textup{G}}\to v_i} \right),
\end{split}
\end{equation*}

\begin{equation*}\small
\begin{split}
&L_{v_i\to c_j^{\textup{G}}}\left[t\right]\\
=&\log \frac{P(\Delta_{i}=t|Y_i=y_i,\{\Se{T}^{l}_{i}\}_{l=1}^{M},\Se{T}^{\textup{G}}_{i}\hspace{-3pt}\setminus\hspace{-3pt} \{c_j^{\textup{G}}\} )}{P(\Delta_{i}=0|Y_i=y_i,\{\Se{T}^{l}_{i}\}_{l=1}^{M},\Se{T}^{\textup{G}}_{i}\hspace{-3pt}\setminus\hspace{-3pt} \{c_j^{\textup{G}}\})}\\
=&\log \frac{P(Y_i=y_i,\{\Se{T}^{l}_{i}\}_{l=1}^{M},\Se{T}^{\textup{G}}_{i}\hspace{-3pt}\setminus\hspace{-3pt} \{c_j^{\textup{G}}\}|\Delta_{i}=t)}{P(Y_i=y_i,\{\Se{T}^{l}_{i}\}_{l=1}^{M},\Se{T}^{\textup{G}}_{i}\hspace{-3pt}\setminus\hspace{-3pt} \{c_j^{\textup{G}}\}|\Delta_{i}=0)}\\
=&\log \frac{P(Y_i=y_i,\Delta_{i}=t)}{P(Y_i=y_i,\Delta_{i}=0)}+\log \frac{P(\{\Se{T}^{l}_{i}\}_{l=1}^{M},\Se{T}^{\textup{G}}_{i}\hspace{-3pt}\setminus\hspace{-3pt} \{c_j^{\textup{G}}\}|\Delta_{i}=t)}{P(\{\Se{T}^{l}_{i}\}_{l=1}^{M},\Se{T}^{\textup{G}}_{i}\hspace{-3pt}\setminus\hspace{-3pt} \{c_j^{\textup{G}}\}|\Delta_{i}=0)}\\
=&\log \frac{P(\Delta_{i}=t|Y_i=y_i)}{P(\Delta_{i}=0|Y_i=y_i)}+\log \frac{P(\{\Se{T}^{l}_{i}\}_{l=1}^{M},\Se{T}^{\textup{G}}_{i}\hspace{-3pt}\setminus\hspace{-3pt} \{c_j^{\textup{G}}\}|\Delta_{i}=t)}{P(\{\Se{T}^{l}_{i}\}_{l=1}^{M},\Se{T}^{\textup{G}}_{i}\hspace{-3pt}\setminus\hspace{-3pt} \{c_j^{\textup{G}}\}|\Delta_{i}=0)}\\
=&L^{\textup{ch}}_{i}\left[t\right]+\log \frac{P(\{\Se{T}^{l}_{i}\}_{l=1}^{M}|\Delta_{i}=t)}{P(\{\Se{T}^{l}_{i}\}_{l=1}^{M} |\Delta_{i}=0)}+\log \frac{P(\Se{T}^{\textup{G}}_{i}\hspace{-3pt}\setminus\hspace{-3pt} \{c_j^{\textup{G}}\} |\Delta_{i}=t)}{P(\Se{T}^{\textup{G}}_{i}\hspace{-3pt}\setminus\hspace{-3pt} \{c_j^{\textup{G}}\} |\Delta_{i}=0)}\\
=&L^{\textup{ch}}_{i}\left[t\right]+\sum\limits_{k\in \Se{T}^{\textup{G}}_{i}\hspace{-1pt}\setminus\hspace{-1pt} \{c_j^{\textup{G}}\}} \frac{P(c_k^{\textup{G}} |\Delta_{i}=t)}{P(c_k^{\textup{G}} |\Delta_{i}=0)}+\sum_{l=1}^{M} \log \frac{P(\Delta_{i,l}=t_l,\Se{T}^{l}_{i})}{P(\Delta_{i,l}=0,\Se{T}^{l}_{i})}\\
=&L^{\textup{ch}}_{i}\left[t\right]+\sum\limits_{k\in \Se{T}^{\textup{G}}_{i}\hspace{-1pt}\setminus\hspace{-1pt} \{c_j^{\textup{G}}\}} L_{ c_k^{\textup{G}}\to v_i}\left[t\right]-\sum\limits_{l=1}^{M}\sum\limits_{k\in \Se{T}_i^l} \log\frac{P(c_k^l|\Delta_{i,l}=0)}{P(c_k^l|\Delta_{i,l}=t_l)} \\
=&L^{\textup{ch}}_{i}\left[t\right]+\sum\limits_{k\in \Se{T}^{\textup{G}}_{i}\hspace{-1pt}\setminus\hspace{-1pt} \{c_j^{\textup{G}}\}} L_{ c_k^{\textup{G}}\to v_i}\left[t\right]-\sum\limits_{l=1}^{M}\Db{I}(t_l=1)\sum\limits_{k\in \Se{T}_i^l} L_{c_k^l\to v_i}.\\
\end{split}
\end{equation*}

Similarly, we derive the CN-to-VN updating laws as follows,
\begin{equation*}\small
\begin{split}
&L_{c_j^l\to v_i}\\
=&\log \frac{P(\Delta_{i,l}=0|c_j^l)}{P(\Delta_{i,l}=1|c_j^l)}=\log \frac{P(c_j^l|\Delta_{i,l}=0)}{P(c_j^l|\Delta_{i,l}=1)}\\
=&\log \frac{P(\sum_{k\in\Se{N}_j^l\hspace{-1pt}\setminus\hspace{-1pt}\{i\}}\Delta_{k,l}=s_j^l)}{P(\sum_{k\in\Se{N}_j^l\hspace{-1pt}\setminus\hspace{-1pt}\{i\}}\Delta_{k,l}=\bar{s_j^l})}\\
=&(1-2s^l_j)2\tanh^{-1}\SB{\prod\nolimits_{k\in \Se{N}_j^l\hspace{-1pt}\setminus\hspace{-1pt}\{i\}} \tanh \left(\frac{L_{v_{k}\to c^l_j}}{2}\right)}\\
=&(1-2s^l_j)\prod\limits_{k\in \Se{N}_j^l\hspace{-1pt}\setminus\hspace{-1pt}\{i\}}\textup{sgn}\left(L_{v_{k}\to c^l_j}\right)\phi\left(\sum\limits_{k\in \Se{N}_j^l\hspace{-1pt}\setminus\hspace{-1pt}\{i\}} \phi\left(\lvert L_{v_{k}\to c^l_j}\rvert\right)\right)\\
=&(1-2s^l_j)\mathcal{L}\left(\phi,\{ L_{v_{k}\to c^l_j} \}_{k\in \Se{N}_j^l\hspace{-1pt}\setminus\hspace{-1pt}\{i\}} \right),
\end{split}
\end{equation*}

\begin{equation*}\small
\begin{split}
&L_{c_j^{\textup{G}}\to v_i}\left[t\right]\\
=&\log \frac{P(\Delta_{i}=t|c_j^{\textup{G}})}{P(\Delta_{i}=0|c_j^{\textup{G}})}=\log \frac{P(c_j^{\textup{G}}|\Delta_{i}=t)}{P(c_j^{\textup{G}}|\Delta_{i}=0)}\\
=&\log \frac{P(\sum_{k\in\Se{N}^{\textup{G}}_j\hspace{-1pt}\setminus\hspace{-1pt}\{i\}} \bold{H}_{j,k}\Delta_k=s_j^{\textup{G}}+\bold{H}_{j,i}t)}{P(\sum_{k\in\Se{N}^{\textup{G}}_j\hspace{-1pt}\setminus\hspace{-1pt}\{i\}} \bold{H}_{j,k}\Delta_k=s_j^{\textup{G}})}\\
=& L'_{c_j^{\textup{G}}\to v_i}\left[s_j^{\textup{G}}+\bold{H}_{j,i}t\right]-L'_{c_j^{\textup{G}}\to v_i}\left[s_j^{\textup{G}}\right].\\
\end{split}
\end{equation*}


\end{proof}

\begin{rem} 
Observe that the first two terms of $L_{v_i\to c_j^l}$ and $L_{v_i\to c_j^{\textup{G}}}\left[t\right]$ represent information flow between neighboring local CNs, and that between neighboring global CNs of $v_i$, respectively, whereas the last term of $L_{v_i\to c_j^l}$ and $L_{v_i\to c_j^{\textup{G}}}\left[t\right]$ denotes information flow between local CNs and global CNs that goes through their shared neighbor $v_i$. This information flow enables the mutual-modification of priors of local CNs and global CNs. This interactive-modification results in decoding error rates better than that of the independent coding schemes.
\end{rem}

\section{Software Implementation}
\label{section: simulation results}
In this section, we first discuss the latency issues of the iterative decoder and present an algorithm that exploits the parallel structure of the decoder for lower latency. Then, we show simulation results of our joint local-global codes with different local vs. global CNs split.

\subsection{Decoding Algorithm}

\begin{algorithm}
\caption{Joint Local-Global Belief Propagation Decoder}\label{algo: Decode2}
\begin{algorithmic}[1]
\Require 
\Statex $Y$: received sequence at Bob's side;
\Statex $R=\bold{H}X$: auxiliary sequence sent from Alice;
\Statex $\bold{H}$: parity-check matrix;
\Statex $\bold{P}$: channel statistics: table of transitional probabilities;
\Statex $r$: number of iterations; 
\Ensure
\Statex $\hat{X}$: estimation of the sequence at Alice's side;
\Statex $L^{\textup{ch}}$: LLR of channel statistics;
\Statex $L^{\textup{cv}}$: LLR of CNs-to-VNs;
\Statex $L^{\textup{vc}}$: LLR of VNs-to-CNs;
\State Compute the syndrome $S=\bold{H}Y-R$;
\For{$i\in V$}
\For{$t\in \textup{GF}(2^M)$} $L^{\textup{ch}}_{i,t}\gets \log (p_{y_i+t,y_i})-\log( p_{y_i,y_i})$;
\EndFor
\For{$1\leq l\leq M$} $L^{\textup{ch}}_{i}(l)\gets f_{M,l}(L^{\textup{ch}}_i)$;
\EndFor
\EndFor
\For{$1\leq m \leq r$}
\For{$i\in V$} $L_{v_i}^{\textup{G}}=\sum\nolimits_{k\in \Se{T}_i^{\textup{G}}} L_{c_k^{\textup{G}}\to v_i}$;
\For{$1\leq l\leq M$} $L_{v_i,l}^{\textup{GL}}=f_{M,l} \left(L_{v_i}^{\textup{G}}\right)$;
\State $L_{v_i}^{l}=\sum\nolimits_{k\in\Se{T}_i^{l}}L_{c_k^l\to v_i}$;
\EndFor
\For{$t\in\textup{GF}(2^M)$} $L_{v_i}^{\textup{LG}}\left[t\right]=\sum\nolimits_{l=1}^{M}\Db{I}(t_l=1)L_{v_i}^{l}$;
\EndFor
\State $L_i=L_i^{\textup{ch}}+L^{\textup{G}}_{v_i}-L_{v_i}^{\textup{LG}}$;
\For{$j\in \Se{T}^{\textup{G}}_i$} $L_{v_i\to c_j^{\textup{G}}}=L_i-L_{c_j^{\textup{G}}\to v_i}$;
\EndFor
\For{$1\leq l\leq M$} $L_i=L_i^{\textup{ch}}+L^{l}_{v_i}+L_{v_i,l}^{\textup{GL}}$;
\For{$j\in \Se{T}^{l}_i$} $L_{v_i\to c_j^{l}}=L_i-L_{c^l_j\to v_i}$;
\EndFor
\EndFor
\EndFor
\For{$j\in \Se{A}^{\textup{G}}$}
\For{$k \in \Se{N}^{\textup{G}}_j$} $L_{k,j}^{\textup{G}}=\mathcal{F}_W \comp h\left(\bold{H}_{j,k}, L_{v_k\to c^{\textup{G}}_j} \right)$;
\EndFor
\State $L_{c_j}^{\textup{G}}=\mathcal{L}\left(\log x, \big \{L_{k,j}^{\textup{G}}\big \}_{k\in\Se{N}^{\textup{G}}_j\hspace{-1pt}}\right)$;
\For{$i \in \Se{N}^{\textup{G}}_j$}
\State $L'_{c_j^{\textup{G}}\to v_i}=\Se{F}_{W}^{-1}\comp\mathcal{L}'\left(\log x, \left(L_{c_j}^{\textup{G}}, L_{i,j}^{\textup{G}}\right)\right)$;
\EndFor
\For{$t\in\textup{GF}(2^M)$} 
\State $L_{c_j^{\textup{G}}\to v_i}=L'_{c_j^{\textup{G}}\to v_i}\left[s_j^{\textup{G}}+\bold{H}_{j,i}t\right]-L'_{c_j^{\textup{G}}\to v_i}\left[s_j^{\textup{G}}\right]$;
\EndFor
\EndFor
\For{$1\leq l\leq M$}
\For{$j\in \Se{A}^{l}$} $L^{l}_{c_j}=\mathcal{L}\left(\phi,\{ L_{v_{k}\to c^l_j} \}_{k\in \Se{N}_j^l} \right)$;
\For{$i\in \Se{N}^{l}_j$}
\State $L_{c^l_j\to v_i}=(1-2s^l_j)\Se{L}'\left(\phi,\left( L^{l}_{c_j}, L_{v_{i}\to c^l_j}\right)\right)$;
\EndFor
\EndFor
\EndFor
\EndFor

\State $Converge=\mathbf{True}$;
\For{$i\in V$} $t=0$, $L_{v_i}^{\textup{G}}=\sum\nolimits_{k\in \Se{T}_i^{\textup{G}}} L_{c_k^{\textup{G}}\to v_i}$;
\For{$1\leq l\leq M$} $L_{v_i,l}^{\textup{GL}}=f_{M,l} \left(L_{v_i}^{\textup{G}}\right)$;
\State $L_{v_i}^{l}=\sum\nolimits_{k\in\Se{T}_i^{l}}L_{c_k^l\to v_i}$;
\If{$L_i^{\textup{ch}}+L^{l}_{v_i}+L_{v_i,l}^{\textup{GL}}<0$} $t_l=1$;
\EndIf
\EndFor
\For{$t\in\textup{GF}(2^M)$} $L_{v_i}^{\textup{LG}}\left[t\right]=\sum\nolimits_{l=1}^{M}\Db{I}(t_l=1)L_{v_i}^{l}$;
\EndFor
\State $L^{\textup{G}}_i=L_i^{\textup{ch}}+L^{\textup{G}}_{v_i}-L_{v_i}^{\textup{LG}}$;
\State $t^{\textup{opt}}=\max\nolimits_{t\in \textup{GF}(2^M)}L^{\textup{G}}_i\left[t\right]$;
\If{$t^{\textup{opt}}\neq t$} $Converge=\mathbf{False}$; $\mathbf{break}$;
\ElsIf{$t^{\textup{opt}}= t$} $\hat{X}_i=t^{\textup{opt}}$;
\EndIf
\EndFor
\State \textbf{Return} $\hat{X}$;
\end{algorithmic}
\end{algorithm}

To make the joint local-global BP algorithm more time-efficient, we apply the following optimization to enable a highly parallel fashioned implementation of the decoder, which is presented as \Cref{algo: Decode2}.
\begin{enumerate}
\item Pre-calculation and reuse of common components among LLRs of different edges in the Tanner graph that connects to the same VN or CN.  
\item Iterative algorithm for computing $f_{M,l}$.\\ To compute $f_{M,l}$, $1\leq l\leq M$, an iterative algorithm that requires $(2^{M+1}-M-2)$ operations can be applied while simply implementing the matrix multiplication requires $M2^M$ operations. Given that $M$ is typically small, i.e., $M\leq 10$, the gain $O(1/M)$ in terms of complexity seems not to be too attractive. However, the iterative algorithm allows a parallel implementation with $M$ layers and requires only XOR and bit-shifting operations, resulting in an nontrivial $O(M/2^M)$ reduction in terms of latency, which implies that our discussion of the iterative algorithm is still efficient and necessary in practical implementation.
\item Property of function $\Se{L}$.\\
Define $\Se{L}'(f,(a,b))=\textup{sgn}(a)\textup{sgn}(b)f^{-1}\left(f(|a|)-f(|b|)\right)$. Then, it is not hard to verify that functions $\Se{L}$ and $\Se{L}'$ have the following relation for all function $f$, set $A$ and $a\in A$:
\begin{equation}
\begin{split}
\Se{L}(f,A\setminus\{a\})=\Se{L}'\left(f,\left(\Se{L}(f,A),a\right)\right).
\end{split}
\end{equation}
By this property, the CNs-to-VNs updating phase can be implemented in a parallel fashion to reduce both complexity and latency.
\end{enumerate}

\subsection{Simulation Results}

\begin{figure}
  \centering
  \includegraphics[width=0.89\linewidth]{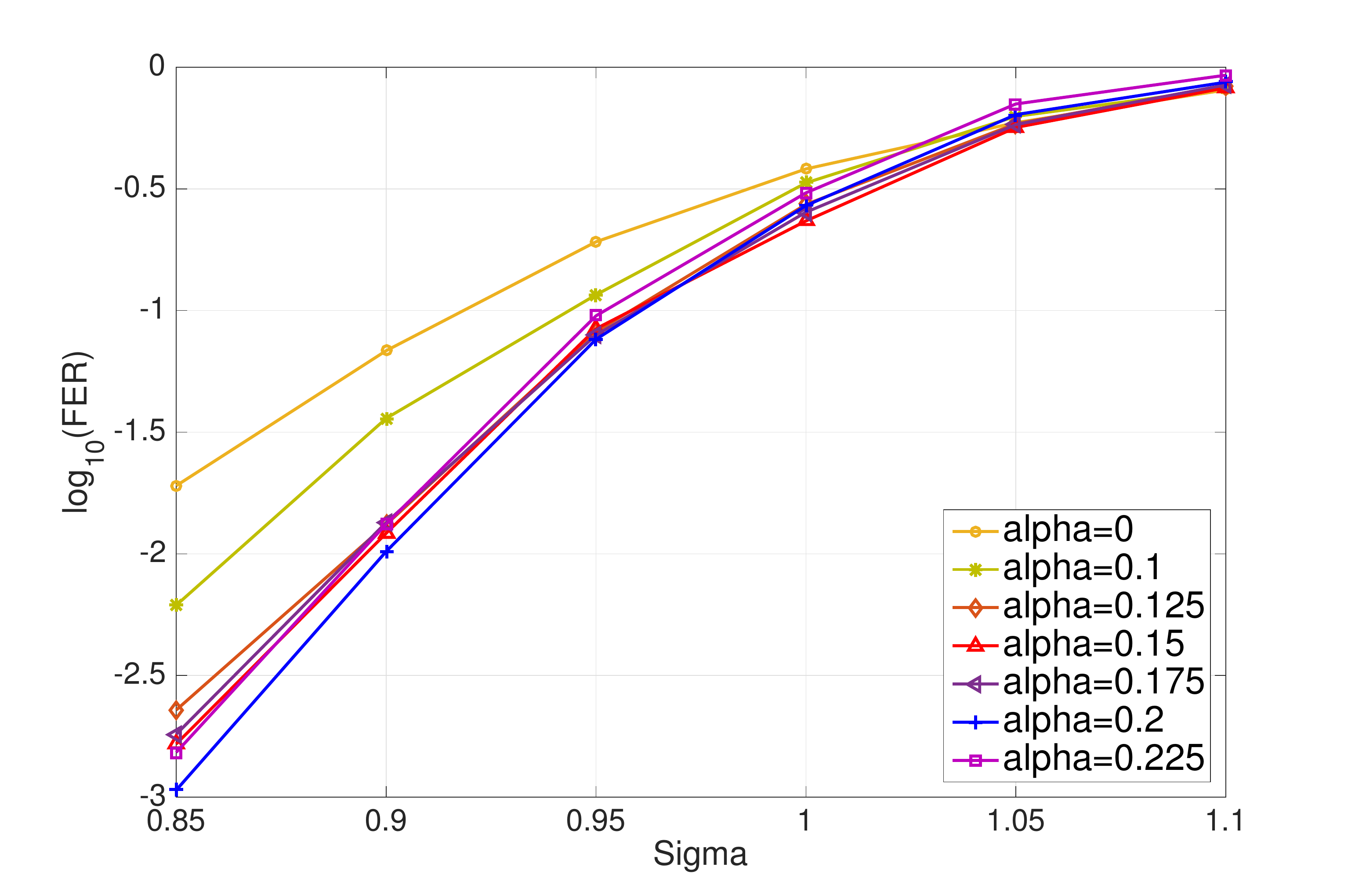}
  \caption{ Simulation results of joint local-global LDPC codes with $M=W=5$, $N=2000$, $R=1/2$, over the channel with $\beta=0$ and $50$ iterations.}
  \label{fig:jointsim}
\end{figure}

Our preliminary investigation suggests that there is a non-trivial dependency between the best local vs. global CNs split and channel conditions. Fig.~\ref{fig:jointsim} presents simulation results of a rate $1/2$ joint local-global codes with length $2000$ and $M=W=5$, for the timing-jitter only channel described as: $P_{Y|X}(y|x)=ce^{-\frac{|y-x|^2}{\sigma}}$ ($c$ is the normalization constant). The local binary codes are randomly constructed as regular LDPC codes with VN degree being $3$, and the global non-binary codes only have degree $1$ VNs. Parameter $\alpha$ denotes the fraction of CNs that are Global CNs, i.e., 
\begin{equation}
\alpha=|\mathcal{A}^{\textup{G}}|/\left(|\mathcal{A}^{\textup{G}}|+|\mathcal{A}^{\textup{L}}|/W\right).
\end{equation} 

As is shown in Fig.~\ref{fig:jointsim}, we notice that by replacing up to $20\%$ of the local binary CNs with global non-binary CNs, the FER is improved by $0.5dB$ in the threshold region and even more in the early waterfall region, which is a significant improvement of the performance without a huge penalty in decoder complexity.

\begin{figure}
  \centering
  \includegraphics[width=0.89\linewidth]{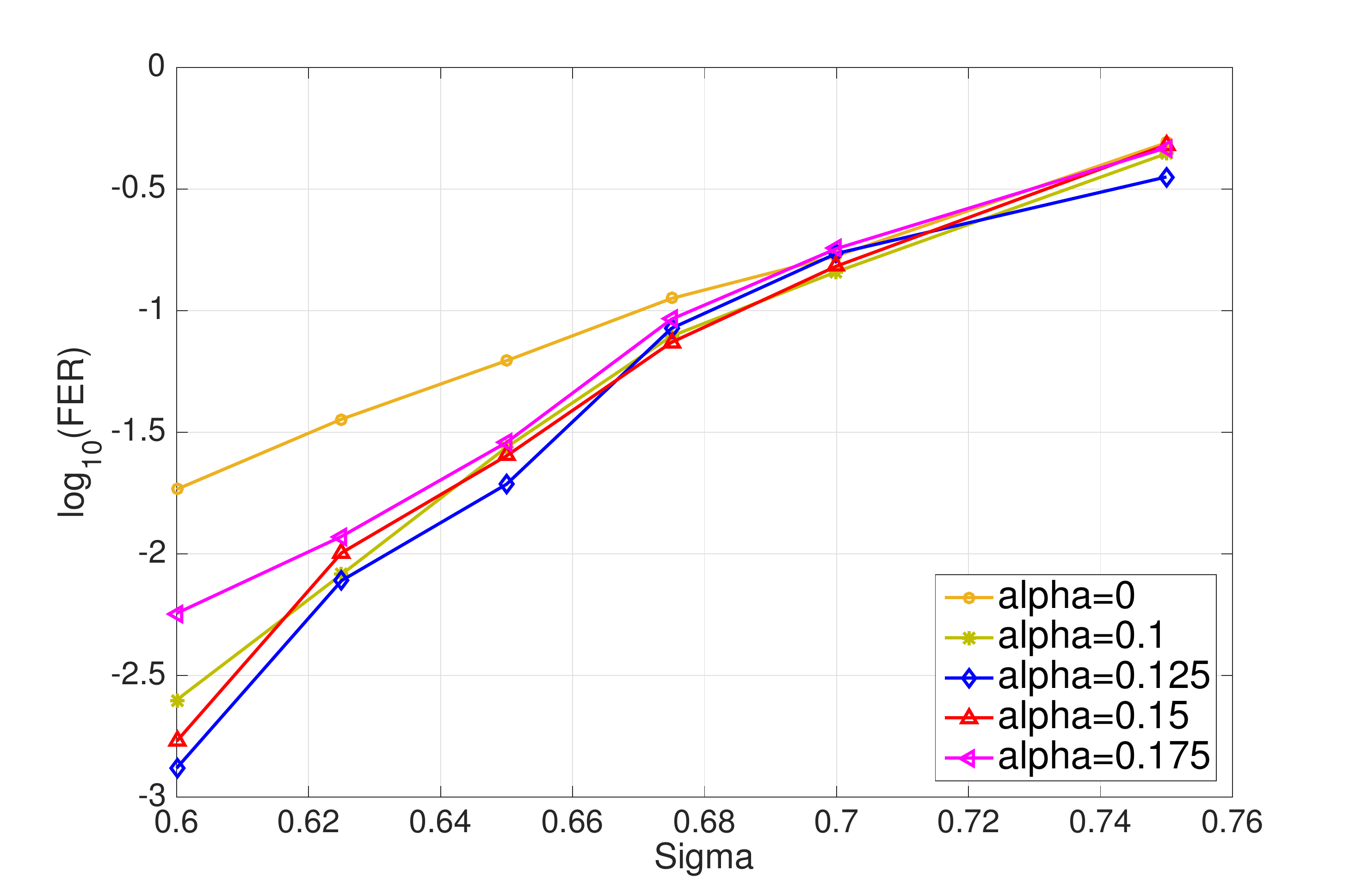}
  \caption{ Simulation results of joint local-global LDPC codes with $M=W=5$, $N=2000$, $R=1/2$ over the joint channel with $\beta=0.001$ and $50$ iterations.}
  \label{fig:jointsim2}
\end{figure}

Among the curves considered, we observe that a higher $\alpha$ does not necessarily result in a lower FER (see e.g., $\alpha=0.2$ vs. $\alpha=0.225$). Moreover, when we add small uniform noise, parameterized by $\beta$, to get the extended channel model $P_{Y|X}(y|x)=ce^{-\frac{|y-x|^2}{\sigma}}+\beta $, we have the same finding: best $\alpha$ was not the highest. Fig.~\ref{fig:jointsim2} presents simulation results of codes having parameters identical to that of Fig.~\ref{fig:jointsim}, on a joint channel with $\beta=0.001$. 
 These examples suggest the existence of the optimal $\alpha$ for the given channel conditions, and our future goal will be to characterize this value.

\section{Conclusion}
\label{section: conclusion}
In this paper, we investigated the channel model in the information reconciliation phase for QKD. To correct errors in this channel, we proposed a code construction that jointly corrects the local errors and the global errors in the QKD channel. To better utilize this coding method, we also presented a balanced modulation scheme that is appropriate for this construction. Simulation results indicate a nontrivial improvement of our proposed scheme over the independent coding scheme. Future work includes structural construction of constituent codes to further improve the performance and the encoding complexity.


\section*{Acknowledgment}
This work has received funding from NSF under the grant No. CCF-BSF 1718389 and No. EFRI 1741707.

\bibliography{ref}

\begin{thebibliography}{10}
\providecommand{\url}[1]{#1}
\csname url@samestyle\endcsname
\providecommand{\newblock}{\relax}
\providecommand{\bibinfo}[2]{#2}
\providecommand{\BIBentrySTDinterwordspacing}{\spaceskip=0pt\relax}
\providecommand{\BIBentryALTinterwordstretchfactor}{4}
\providecommand{\BIBentryALTinterwordspacing}{\spaceskip=\fontdimen2\font plus
\BIBentryALTinterwordstretchfactor\fontdimen3\font minus
  \fontdimen4\font\relax}
\providecommand{\BIBforeignlanguage}[2]{{%
\expandafter\ifx\csname l@#1\endcsname\relax
\typeout{** WARNING: IEEEtran.bst: No hyphenation pattern has been}%
\typeout{** loaded for the language `#1'. Using the pattern for}%
\typeout{** the default language instead.}%
\else
\language=\csname l@#1\endcsname
\fi
#2}}
\providecommand{\BIBdecl}{\relax}
\BIBdecl
\renewcommand{\BIBentryALTinterwordstretchfactor}{4}

\bibitem{Zhong2015a}
T.~Zhong \emph{et~al.}, ``{Photon-efficient quantum key distribution using
  time-energy entanglement with high-dimensional encoding},'' \emph{New Journal
  of Physics}, vol.~17, no.~2, Feb. 2015.

\bibitem{Zhang2013}
Z.~Zhang \emph{et~al.}, ``{Unconditional Security of Time-Energy Entanglement
  Quantum Key Distribution Using Dual-Basis Interferometry},'' \emph{Physical
  Review Letters}, vol. 112, no.~12, Mar. 2014.

\bibitem{zhou2013layered}
H.~Zhou, L.~Wang, and G.~Wornell, ``Layered schemes for large-alphabet secret
  key distribution,'' in \emph{2013 Information Theory and Applications
  Workshop (ITA)}.\hskip 1em plus 0.5em minus 0.4em\relax IEEE, 2013, pp.
  1--10.

\bibitem{jiang2018high}
X.-Q. Jiang, S.~Yang, P.~Huang \emph{et~al.}, ``High-speed reconciliation for
  {C}{V}{Q}{K}{D} based on spatially coupled {L}{D}{P}{C} codes,'' \emph{IEEE
  Photonics Journal}, vol.~10, no.~4, pp. 1--10, 2018.

\bibitem{martinez2013key}
J.~Martinez-Mateo, D.~Elkouss, and V.~Martin, ``Key reconciliation for high
  performance quantum key distribution,'' \emph{Scientific Reports}, vol.~3,
  2013.

\bibitem{Wehnereaam9288}
S.~Wehner, D.~Elkouss, and R.~Hanson, ``Quantum internet: A vision for the road
  ahead,'' \emph{Science}, vol. 362, no. 6412, 2018.

\bibitem{olmos2015scaling}
P.~M. Olmos and R.~L. Urbanke, ``A scaling law to predict the finite-length
  performance of spatially-coupled {L}{D}{P}{C} codes,'' \emph{IEEE {T}rans.
  {I}nformation {T}heory}, vol.~61, no.~6, pp. 3164--3184, 2015.

\bibitem{declercq2007decoding}
D.~Declercq and M.~Fossorier, ``Decoding algorithms for nonbinary {L}{D}{P}{C}
  codes over {G}{F}(q),'' \emph{IEEE Trans. Communications}, vol.~55, no.~4,
  pp. 633--643, 2007.

\bibitem{hareedy2018spatially}
A.~Hareedy \emph{et~al.}, ``Spatially-coupled code design for partial-response
  channels: optimal object-minimization approach,'' in \emph{2018 IEEE Global
  Communications Conference (GLOBECOM)}.\hskip 1em plus 0.5em minus 0.4em\relax
  IEEE, 2018, pp. 1--7.

\bibitem{esfahanizadeh2018finite}
H.~Esfahanizadeh, A.~Hareedy, and L.~Dolecek, ``Finite-length construction of
  high performance spatially-coupled codes via optimized partitioning and
  lifting,'' \emph{IEEE {T}rans. {C}ommunications}, vol.~67, no.~1, pp. 3--16,
  2018.

\bibitem{richardson2001design}
T.~J. Richardson, M.~A. Shokrollahi, and R.~L. Urbanke, ``Design of
  capacity-approaching irregular low-density parity-check codes,'' \emph{IEEE
  {Trans}. Information Theory}, vol.~47, no.~2, pp. 619--637, 2001.

\end{thebibliography}
\bibliographystyle{IEEEtran}

\end{document}